\documentclass[aps,prb,twocolumn,groupedaddress,showpacs]{revtex4}

\usepackage{graphicx}
\usepackage{float}
 \usepackage{epsfig}
 \usepackage{color}
 \usepackage{soul}
 \usepackage{amsmath}
 \usepackage{amssymb}
\usepackage{amsfonts}
\usepackage{listings}
\usepackage{mathrsfs}
\usepackage{epstopdf}
\begin{document}

\title{Theory of the magnetic and metal-insulator transitions in RNiO$_3$ bulk and layered structures.}

\author{Bayo Lau and Andrew J. Millis}
\affiliation{Department of Physics, Columbia University, 538 West 120th Street, New York, NY, USA 10027}
\date{\today}

\begin{abstract}
A slave rotor-Hartree Fock formalism is presented for studying the properties of the p-d model describing perovskite transition metal oxides, and a  flexible and efficient numerical formalism is developed for its solution. 
The methodology is shown to yield, within an unified formulation, the significant aspects of the rare earth nickelate phase diagram, including the paramagnetic metal state observed for the LaNiO$_3$ and the correct ground-state magnetic order of insulating compounds. It is then used to elucidate ground state changes occurring as morphology is varied from bulk to strained and un-strained thin-film form.
For ultrathin films, epitaxial strain and charge-transfer to the apical out-of-plane oxygen sites  are shown to have significant impact on the phase diagram.
\end{abstract}

\pacs{71.30.+h,73.21.-b,75.25.-j,75.25.Dk}

\maketitle

Understanding the unusual electronic behavior of  transition-metal oxides has been a long-standing question in condensed matter physics,\cite{mit} and interest has   intensified following the demonstration \cite{Ohtomo02} that the materials could be used as components of atomically precise oxide heterostructures. \cite{Mannhart08,Hwang12,Chakhalian12} The theoretical challenge posed by the materials is to treat simultaneously the strong local correlations in the transition metal  d-orbitals and their substantial hybridization with oxygen $p$  orbitals. In some systems, the $p$ orbitals can be integrated out and the physics represented in terms of a (possibly multi band) Hubbard model representing the $d$ orbitals only, for which many theoretical methods are available\cite{mit}. However, in many cases the charge transfer between $p$ and $d$ orbitals is large enough that the $p$ states cannot be neglected. This ``negative charge transfer energy" regime\cite{zsa} has been less extensively studied.
While much useful information has been provided by density functional theory (DFT) \cite{Jones89} and extensions such as  DFT+U\cite{Anisimov93,dft0,dft1,Han11,Han12,Blanca-romero11,Anisimov99}, hybrid functionals \cite{Gou11,Puggioni12}, and DFT+DMFT \cite{Kotliar06,Park12}, these methods are computationally  intensive, so that the large supercells required for long-period ordered phases are difficult to study.
Furthermore,  the variety of experimental bulk and superlattice configurations and of many-body phenomena emphasizes the need for a model-system treatment that encapsulates the essential physics so the importance of different contributions can be disentangled.

\begin{figure}[]
\centering
\includegraphics[trim = 15mm 0mm 10mm 10mm, clip, width=\columnwidth]{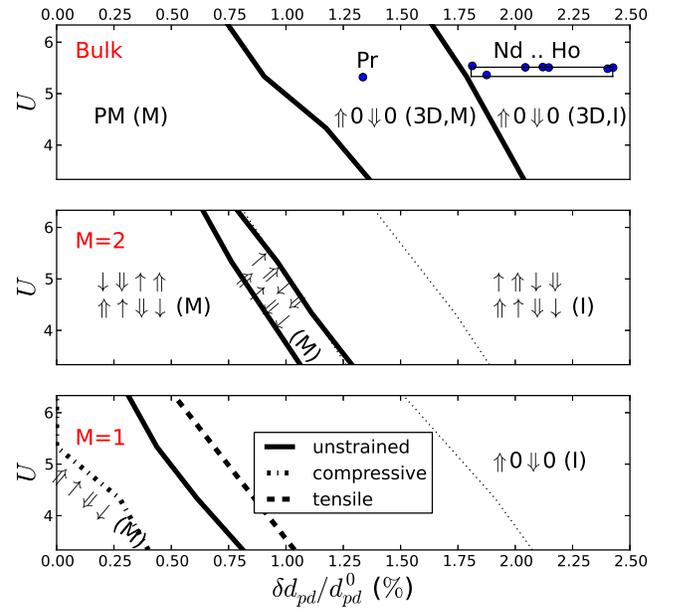}
\caption{\label{fig:summary} (Top to bottom) Bulk, 2-, and 1-layer phase diagram on the $u$-$\delta d_{pd}$ plane.
Slight charge ordering  accompanies the spin ordering for $\delta d_{pd}\neq$0.
$\Uparrow$/$\uparrow$/0 denote large moment/small moment/non-magnetic sites of the long-period magnetic ordering patterns  detailed in the text.  In the lower two panels the light dotted line indicates the phase boundary obtained without the energy shift on the apical oxygen.
}
\end{figure}

The rare earth nickelates\cite{nio3_rev}, $ReNiO_3$, are an important case in point. 
According to standard valence counting arguments, Ni is in the 3+ oxidation state with a $d^7$ valence configuration. However, photoemission experiments \cite{mizokawa} reported that the dominant GS configuration is $d^8\underline{L}$, in agreement with  unrestricted Hartree-Fock (UHF)\cite{mizokawa,hf1} and DFT+U\cite{Blanca-romero11,Han12,Anisimov99}, and DFT+DMFT\cite{Park12} approaches.  Additionally, O 1s x-ray absorption experiments have found significant hole concentration on the oxygen $p$ orbitals\cite{mizokawa,xas0,xas1}.  This large degree of charge transfer places the material near the negative charge-transfer regime\cite{zsa,hf1}.
As  $Re$ is varied across the lanathanum row of the periodic table, the ground state (GS) of bulk materials changes from paramagnetic metal (PM-M) to correlated insulator \cite{mit,nio3_rev,Torrance92}. The correlated insulator phases exhibit a rock-salt-pattern lattice distortion \cite{garcia1,alonso0,alonso1,medarde0,Medarde08} and a nontrivial long-period magnetic ordering \cite{alonso0,alonso1,garcia0,scagnoli0,scagnoli1,bodenthin}. In ultra-thin films a   metal-insulator transition which is  apparently {\em unaccompanied} by rock-salt-pattern lattice distortion occurs as film thickness and strain are varied \cite{liu,liu1,keimer,chakhalian,son0,son1,moon}.  Understanding how these apparently different transitions can occur within a single formulation is an important open theoretical challenge\cite{dft0,dft1,mazin,Gou11,Han11,Han12,Puggioni12,Blanca-romero11,chakhalian,Anisimov99,hf1,balents_prl,balents_prb,Park12}.

In this paper, we develop  a mean-field (MF) approach  based on a combination of slave-rotor (SR)\cite{rotor0,rotor1,rotor2,rotorj0,rotorj1} and Hartree-Fock (HF) methods, along with an efficient and flexible numerical formalism for solving the MF equations.   The method is powerful enough to permit the examination of the large supercells needed to investigate long-period ordering patterns in the context of realistic crystal structures. The results reconcile the bulk and film phase diagrams and allows us to identify the key role played by the oxygen degrees of freedom and the lattice distortions. 

Figure ~\ref{fig:summary} summarizes our key new results. It presents phase diagrams in the space of Ni charging energy $U$ and amplitude $\delta d_{pd}$ of  rocksalt-type lattice distortion (defined more precisely below). The upper panel shows that, in the absence of oxygen breathing distortion, bulk materials are PM-M at any value of $U$, consistent with experiment\cite{nio3_rev}.
As the lattice is distorted, transitions occur, first to a metallic magnetic phase and then to a magnetic insulator. The nontrivial ordering wavevector found in experiment\cite{alonso0,alonso1,garcia0,scagnoli0,scagnoli1,bodenthin} is correctly obtained as a 3D 16-formula-unit magnetic pattern of the type $\Uparrow$0$\Downarrow$0, denoting the z-projection of magnetic moment detailed below. The 4 $\Uparrow$ and 4 $\Downarrow$ Ni sites are surrounded by 6 spin-0 Ni sites, which in our interpretation form  singlets with its 6 oxygen neighbors. Each of the 8 spin-0 Ni sites are connected to 3 $\Uparrow$ and 3 $\Downarrow$ Ni sites via oxygen bonds. States with the $\uparrow\uparrow\downarrow\downarrow$ pattern are unstable against PM or $\Uparrow 0\Downarrow 0$ solutions, and indeed are not observed in experiments\cite{scagnoli0}. In agreement with symmetry-based arguments\cite{balents_prl}, slight charge ordering always accompanies the spin ordering. For reasonable parameter values the change in electronic structure across the $Re$ series is properly accounted for, with one exception: the magnetic metallic phase found in theory is not observed in experiment. This is discussed in more detail below.  
The lower panels show the evolution of the phase diagram with film thickness $M$  and (bottom panel) with applied strain,  revealing that in ultra-thin films an insulating phase can occur even in the absence of rocksalt-type lattice distortions.

Our theoretical approach begins from   a $p$-$d$ lattice model of the form
$H = \sum_{k} H^{pd}_{k}  + \sum_l \left( U_l + J_l\right)$,
with $k$ and $l$ sum over crystal momenta and Ni sites, respectively. The parameters of this model may be obtained from, e.g. maximially localized Wannier fits to DFT results\cite{Marzari97}, but the precise form is not important for this paper (see also Ref.~\onlinecite{Wang11} for another example of the insensitivity of results to the precise p-d model band parameters).  We adopt a simplified scenario in which we retain the Ni $e_g$ and O $2p\sigma$ orbitals, with  nearest-neighbor $p$-$d$ and $p$-$p$ hopping. The difference between the   bare Ni-d and O-2p energies $\varepsilon_d$-$\varepsilon_p$ is important, as discussed below.

The structure of the  ReNiO$_3$ materials is derived from the ideal ABO$_3$ cubic perovskite structure which is a lattice of corner-sharing oxygen octahedra, each containing a Ni site at the center. The crystal structure of the actual materials is distorted from this structure by rotations of the octahedra which are not important for our purposes and,   in the insulating cases, by a two-sublattice distortion in which 
adjacent Ni's have significant different Ni-O bond lengths\cite{garcia1,alonso0,alonso1,Medarde08}. To incorporate the bond disproportionation in the Hamiltonian under the cubic approximation, we scale the hopping according to the Harrison rule\cite{harrison}, $t_{pd}=t_{pd}^0\left(1+\delta d_{pd}/d_{pd}^0\right)^{-4}$ and $t_{pp}=t_{pp}^0\left(1+\delta d_{pp}/d_{pp}^0\right)^{-3}$, with $d_{pd}^0=1.95\AA$ and $d_{pp}^0=\sqrt{2}d_{pd}^0$. 
An additional effect may occur in  layered structures. Liu et al. \cite{liu} showed that the presence of Al at the interface would deplete holes on the out-of-layer oxygen sites linking Al and Ni, raising the charge-transfer energy from those apical sites by $\sim$1eV. We model the $M$-layer 2D structures using supercells with $M$ NiO$_3$ units in the z-direction, terminated on both ends with apical oxygen sites whose $e_p$ are shifted by -1eV. For reference we also performed some calculations without such shift.

The  interaction terms $U_l$ and $J_l$ contain the on-site repulsion and the Hund's interactions in the rotationally-invariant Slater-Kanamori form\cite{mit}.
\begin{equation}
U_l=U\sum_{(m\sigma)>(m'\sigma')} n_{m\sigma}n_{m'\sigma'}
\end{equation}
\begin{eqnarray}
J_l=&-&\frac{4}{3}j\sum_\sigma n_{a\sigma}n_{b\sigma}+\frac{5}{3}j\sum_m n_{m\uparrow}n_{m\downarrow}\nonumber\\
&-&j\sum_\sigma\left( \frac{1}{3}n_{a\sigma}n_{b-\sigma}+d_{a\sigma}^\dag d_{b-\sigma}^\dag d_{b\sigma} d_{a-\sigma}\right)\nonumber\\
&+&j\left( d_{a\uparrow}^\dag d_{a\downarrow}^\dag d_{b\downarrow} d_{b\uparrow} +h.c.\right)
\end{eqnarray}
In this expression, $J_l$ differentiates only states with the same occupancy but inequivalent configurations.

To treat the on-site Coulomb interactions, we adopt the slave rotor (SR)   approach\cite{rotor0,rotor1,rotor2,rotorj0,rotorj1}. The method was originally applied to the Hubbard model and then to other $d$-only models with Hund's-like interactions\cite{rotorj0,rotorj1} under the $j<<U$ approximation. We apply it here to the p-d model, noting that   for RNiO$_3$  j$\sim$1eV is much smaller than either the d-d repulsion and the electron bandwidth.
For each $e_g$ site, the approach introduces an auxiliary SR field, $\theta_l \in [0,2\pi)$, and decomposes electron operators as $d^\dag_{l\alpha\sigma}\rightarrow f^\dag_{l\alpha\sigma}e^{i\theta_l}$. The consistency of SR state and d-occupancy is enforced by the constrain 
\begin{equation}
\widehat{L}_l = \frac{\partial}{i\partial\theta_l}=\sum_{\alpha\sigma}\left(f^\dag_{\alpha\sigma} f_{\alpha\sigma}-\frac{1}{2}\right) \label{eq:constrain}
\end{equation} 
The spectrum of $U_l^{(\theta)}\propto\widehat{L}_l^2$. We treat the large $U_l$ with $\theta$ and the small j-scale $J_l$ with $f^\dag$ using the weak coupling Hartree-Fock approximation. That is, we solve the $p$-$d$ model with the single-Slater-determinant ansatz $|MF\rangle = |p,f\rangle |\theta\rangle$.  We follow previous SR applications using Lagrange multipliers, $h_l$. Along with constraint Eq.~\ref{eq:constrain} and up to a constant, the system of equations reads
\begin{eqnarray}
H_{p,f}= &&  \sum_l J^{(f)}_l+\sum_{l\alpha\sigma}(\epsilon_d-\epsilon_p+\frac{3}{2}U-h_l) f^\dag_{l\alpha\sigma}f_{l\alpha\sigma}\nonumber\\
&+&\sum_{l\epsilon\alpha\sigma}\langle e^{-i\theta_l}\rangle V_{\epsilon\alpha} p^\dag_{l+\epsilon\sigma}f_{l\alpha\sigma}+h.c.\nonumber\\
&+&\sum_{l\epsilon\delta\sigma} t_{pp,\epsilon,\delta}p^\dag_{l+\epsilon\sigma}p_{l+\epsilon+\delta,\sigma}+h.c.\label{eq:HF}
\end{eqnarray}
\begin{eqnarray}
H_\theta=&&\sum_l \frac{U}{2}\widehat{L}_l^2+h_l\widehat{L}_l\nonumber\\
&+&\sum_l e^{i\theta_l}\sum_{\alpha\epsilon\sigma}V_{\epsilon\alpha}\langle p^\dag_{l+\epsilon\sigma}f_{l\sigma}\rangle + h.c.\label{eq:rotor}
\end{eqnarray}
where $V^0_{\epsilon\alpha}$ is given by the product of $t^0_{pd}$ and the Slater-Koster orbital symmetry factors in Table I of Ref.~\onlinecite{Slater54}.

We solve the MF equations without any restriction other than the single-Slater-determinant
assumption. A 3D Bravais lattice of Ni$_{16}$O$_{48}$ supercells is required for unrestricted modeling of the $(1/2,0,1/2)$ pattern with respect to the orthorhombic unit cell. To systematically capture the effects of dimensionality, we model the $M$-layer structures  using a 2D Bravais lattice of Ni$_{4M}$O$_{12M+4}$ supercells, which is connected to the 3D Ni$_{16}$O$_{48}$ supercell for $M\rightarrow\infty$. The system is solved by the standard T=0 iterative procedure for up to 65536 supercells. Without much optimization, the worst case initial condition with bi-partite charge-, orbital-, and magnetic-order would converge into a PM-M or $(1/2,0,1/2)$-ordered insulator within 30 cpu hours on an Opteron-2350 cluster.

We perform bulk calculations for (u,$e_d$) such that $\langle n_{e_g}\rangle\sim 2$ as found in UHF with parameters fitted to photoemission spectrum\cite{hf1}, DFT\cite{Blanca-romero11,Han12,Anisimov99}, and DMFT+DFT\cite{Park12}. For the reference set of $U=5.3$, $j=1$, $t_{p_\sigma,d_{x^2-y^2}}=1.5$, $t_{pp}=0.5$, $\Delta_{pd}=e_d-e_p+U=0.3$, the undistorted 3D lattice has $\langle n_{e_g}\rangle=1.95$, slightly less than those of the above calculations. To isolate dimensionality effects, we use the bulk's ($u$,$e_d$) pairs, in addition to the aforementioned apical $e_p$ shift, for layered calculations.

\begin{figure}[]
\centering
\includegraphics[trim = 12mm 0mm 5mm 15mm, clip, width=1\columnwidth]{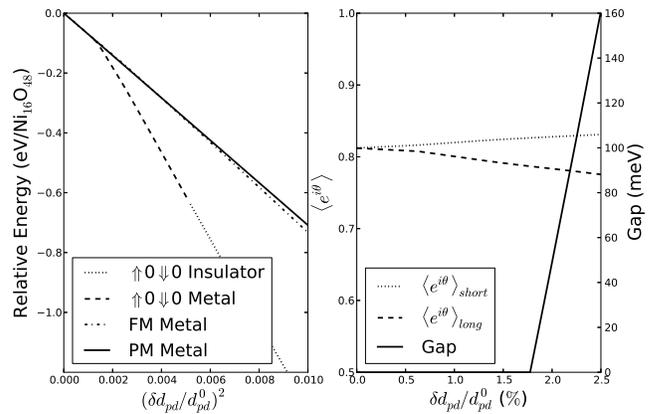}
\caption{\label{fig:bulk_misc} (Left) Stable solutions' MF energy. (Right) rotor renormalization and insulating gap for the bulk system at $U=5.3$.}
\end{figure}

The left panel of Fig.~\ref{fig:bulk_misc} shows representative results for the energies of different locally stable phases computed at interaction (U,j)=(5.3,1) as a function of $(\delta d_{pd})^2$. We see that in agreement with experiment, the undistorted lattice is a PM-M. As oxygen breathing disproportionation $\delta d_{pd}$ is increased a transition to a period-4 metallic magnetic state occurs. At yet larger distortion a metal-insulator transition occurs.  The ferromagnetic states found in UHF \cite{mizokawa,hf1}, DFT+U \cite{Blanca-romero11,Han12,Anisimov99} and DFT+DMFT \cite{Park12} become stable only at larger $\delta d_{pd}$ and gain less energy than does the experimentally observed ordering pattern. Also, $E$$\sim$$-\delta d_{pd}^2$ implies an equilibrium distortion ultimately determined by the aharmonic lattice restoration force beyond the scope of this study.

The dashed and dotted lines in the right panel of Fig.~\ref{fig:bulk_misc} are the SR renormalizations  of the p-d hopping (Eq.~\ref{eq:HF}) on the two sublattices of the distorted structure. We see that the Coulomb renormalizations are not large,
and are only weakly dependent on sublattice. This renormalization has a moderate effect on physical properties, e.g.
 Fermi velocity renormalization at (U,j)=(5.3,1) is $v_{5.3,1}/v_{0,0}\sim$0.65.
The relatively modest SR renormalizations confirms that near the ``negative charge transfer'' region of the phase diagram \cite{zsa} the on-site Coulomb correlations do not drive the phase transition in an orthodox way, in particular the physics is far from the $\langle e^{i\theta}\rangle\rightarrow 0$ Brinkman-Rice insulator.
Even then,  the renormalized kinetic terms, combined with lattice distortion and the magnetic order opens a gap in the density of states. We stress that SR renormalization $\langle e^{i\theta}\rangle$ due to $U$ is required for insulation within physical parameter range.

Parameters scaled for different Re bulk materials\cite{Medarde08,garcia1,alonso0,alonso1} are marked on the phase diagram in top panel of Fig.~\ref{fig:summary}. It captures all experimentally observed phases over the Re series, with the exception of Re=Pr, which is predicted to be a metal instead of an insulator, albeit with the correct charge and magnetic order.  This can be understood by noting that our approach underestimates the insulating gap, which is  160meV at $\frac{\delta d_{pd}}{d_{pd}^0}$=2.5\% compared to $\sim$200meV in a 2-site DFT+DMFT study\cite{Park12}, and that Re=Pr is also extrapolated to be a metal from those Re=Lu results.

In agreement with experiments, the $\Uparrow$0$\Downarrow$0 insulator (top panel of Fig.~\ref{fig:summary}) has a (1/2,0,1/2) magnetic structure with respect to the orthorhombic unit cell, a rock-salt charge ordering pattern, but without orbital ordering. Let us identify the Ni sites with longer/shorter Ni-O bond length as Ni$_{\mbox{long}}$ and Ni$_{\mbox{short}}$ and consider the $\frac{\delta d_{pd}}{d_{pd}^0}$=2.5\% solution. The occupancy at those sites are $n_{\mbox{long}}=2.03$ and $n_{\mbox{short}}$=1.90 . This disproportionation can be understood by measuring the Ni-O hybridization order parameter
$\widehat{O}_{l}=\sum_{\epsilon\alpha\sigma} V^0_{l\epsilon\alpha} e^{-i\theta_l} p^\dag_{l+\epsilon\sigma}f_{l\alpha\sigma}+h.c.$
We found that $\frac{\langle O_{\mbox{short}}\rangle}{\langle O_{\mbox{long}}\rangle}=1.4$, which is greater than expected from the corresponding $\delta t_{pd}/t_{pd}^0\sim $10\%  hopping modulation. This shows that Ni$_{\mbox{short}}$ is strongly hybridized with the oxygen bands while Ni$_{\mbox{long}}$ retains its d$^8$ characteristic. The magnetic moments are $m_{\mbox{long}}$=1.1, which agrees with experiment\cite{alonso0,alonso1}, and $m_{\mbox{short}}$=0. For charge-transfer insulators, the effective physics of strong p-d hybridization is a strong antiferromagnetic $\overline{S}_d\cdot\overline{S}_p$ spin correlation. In this context, having $m_{\mbox{short}}=0$ in the single-Slater-determinant approximation is evident of a d$^8$-high-spin state forming a singlet with its oxygen neighbors. The formation of $p$-$d$ singlets has also been reported in a two-site DFT+DMFT study which predicted FM GS\cite{Park12}.

\begin{figure}[]
\centering
\includegraphics[trim = 10mm 1mm 20mm 15mm, clip, width=\columnwidth]{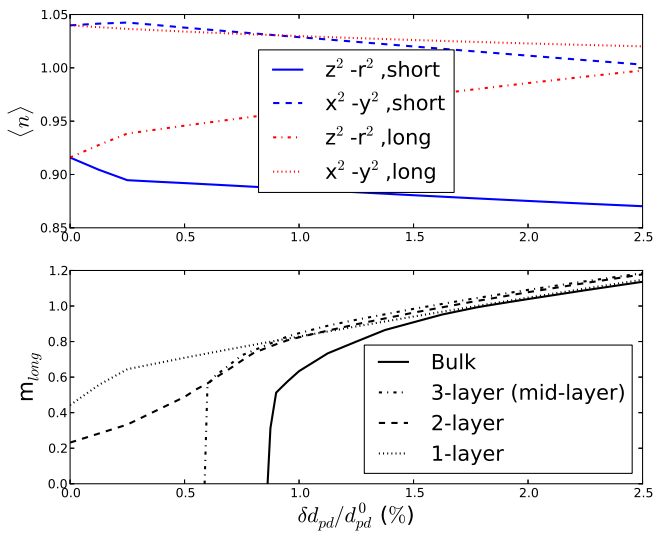}
\caption{\label{fig:misc_vs_x_layer} Results for $u=7$. (Top) $d$-occupancy for the 1-layer structure. (Bottom) $m_{long}$ magnetic moment for bulk, middle layer of 3-layer, 2- and 1-layer structures.}
\end{figure}

We now discuss the effects of dimensional confinement. The bulk and $M\geq 3$ layers structures have similar phase diagrams, with increasing ordered and insulating region for decreasing $M$ and increasing $U$. In particular, the lower panel of Fig.~\ref{fig:misc_vs_x_layer} shows that PM-M exists for $M\geq 3$. The phase diagram of 2-, and 1-layer ultra-thin films are shown in the lower panels of Fig~\ref{fig:summary}. We see that, for $M\leq 2$ layers, the ground state is always magnetically ordered, even in the absence of the bond disproportionation. Further,  the distortion amplitude  required for the transition to the insulating phase decreases with increasing $u$ and decreasing $M$. At (u,j)=(5.3,1), the metal-insulator boundary $(\frac{\delta d_{pd}}{d_{pd}^0})_{MIT}$ is reduced from 1.9\% to 0.96\% and 0.44\% for 2- and 1-layer, respectively. A comparison between the solid and dotted lines in Fig.~\ref{fig:summary} shows that charge transfer to the out-of-plane orbitals\cite{liu} greatly enlarge the insulating regions.

To study the effects of epitaxial strain, we focus on 1-layer structure for concreteness. We include the strain effects by scaling hopping integrals according to the Harrison rule\cite{harrison} for the geometry specified in Ref.~\onlinecite{keimer}, which introduced compressive (tensile) strain with LaSrAlO$_4$ (SrTiO$_3$) such that the lattice parameters are a=b=3.769 (3.853)\AA~  and c = 3.853 (3.790)\AA~. Breathing-mode distortion is then introduced as discussed above. In agreement with experiments on ultra-thin layers\cite{keimer,liu1,moon}, Fig.~\ref{fig:summary} shows that compressive (tensile) epitaxial strain enlarges (shrinks) the insulating region. Under realistic values of $U$ and compressive stain, ordered insulator is possible for $\delta d_{pd}$=0, agreeing with experiments which found breathing-mode distortion only under tensile strain\cite{chakhalian}. Note that further shift in apical oxygen $e_p$ also decreases $\delta d_{pd}$ required for insulation.

Figure \ref{fig:misc_vs_x_layer} demonstrates an orbital polarization of $\sim$5\%. We also found that the polarization decreases (increases) with compressive (tensile) strain, in agreement with a DFT+U study\cite{Han12,Han11}. For example, LaSrAlO$_4$ (SrTiO$_3$) substrate changes the polarization to $\sim$4 (6)\%. While DFT+U predicted FM order, our $M$=1,2 GS have an AFM in-plane order of $\Uparrow\uparrow\Downarrow\downarrow$, which differs slightly from the bulk's  $\Uparrow$0$\Downarrow$0 pattern by having small $m_{short}$$\neq$0 which decreases with increasing distortion.
The 2-layer structure has an additional magnetic transition between FM and AFM ordering in the z-direction.

Lastly, we compare our results with other theoretical works. In our formulation, the bulk PM-M develops into $\Uparrow$0$\Downarrow$0-I for $\frac{\delta d_{pd}}{d_{pd}^0}>$1.9\% compared to UHF's $\frac{\delta d_{pd}}{d_{pd}^0}>$7.5\% transition from ferromagnetic to MCO state with similar pattern\cite{hf1}. The pattern is also similar to the $j/U>1$ S-SDW state in the $d$-only model\cite{balents_prb}, but we have  $j/U<1/5$ and a bond and charge disproportionation with $d^{8+\delta}\underline{L}d^{8-\delta}\underline{L}$ involving strong $pd$ hybridization. Unlike the $d$-only model, our results are sensitive to dimensional confinement.
Even though different DFT implementations have been employed to capture the bulk LaNiO$_3$ PM-M\cite{Gou11} and LaNiO$_3$/LaAlO$_3$ ordered insulating layer\cite{Puggioni12}, the demonstration of all bulk and layered phases by a single formulation had been elusive to the best of our knowledge.

Our formulation bridges the gap between HF-like approaches and the expensive cluster DMFT. The results connect all ReNiO$_3$ delicate phases in bulk and layer form and also provide detailed insights about them. This demonstrates a viable pathway of treating systems with different superstructures as well as compounds such as Fe and Co oxides with strong $p$-$d$ hybridization and large number of partially filled strongly correlated orbitals. For example, Sr$_2$FeO$_4$ has also been shown to exhibit strong hybridization and non-trivial magnetic order\cite{Rozenberg98}.

We thank G. A. Sawatzky, H. Chen, H. T. Dang, R. Fernandez, S. Park, and D. Zgid for helpful discussions. This effort is supported by US National Science Foundation under grant NSF-DMR-1006282.

\end{document}